\documentclass[%
reprint,
superscriptaddress,
amsmath,amssymb,
aps,
pra
]{revtex4-1}

\usepackage{graphicx}
\usepackage{dcolumn}
\usepackage{bm}
\usepackage{physics}

\begin{document}


\title{Approaching ultra-strong coupling in Transmon circuit-QED\\ using a high-impedance resonator}

\author{Sal J. Bosman}
\thanks{These authors contributed equally to this manuscript.}
\affiliation{%
Kavli Institute of NanoScience, Delft University of Technology,\\
PO Box 5046, 2600 GA, Delft, The Netherlands.
}%

\author{Mario F. Gely}
\thanks{These authors contributed equally to this manuscript.}
\affiliation{%
Kavli Institute of NanoScience, Delft University of Technology,\\
PO Box 5046, 2600 GA, Delft, The Netherlands.
}%

\author{Vibhor Singh}%
 \affiliation{%
Department of Physics, Indian Institute of Science, Bangalore 560012, India
}%

\author{Daniel Bothner}%
 \affiliation{%
Kavli Institute of NanoScience, Delft University of Technology,\\
PO Box 5046, 2600 GA, Delft, The Netherlands.
}%

\author{Andres Castellanos-Gomez}
\affiliation{%
 Instituto de Ciencia de Materiales de Madrid, CSIC, Madrid 28049, Spain}

\author{Gary A. Steele}

\affiliation{%
Kavli Institute of NanoScience, Delft University of Technology,\\
PO Box 5046, 2600 GA, Delft, The Netherlands.
}%

\date{\today}

\begin{abstract}

In this experiment, we couple a superconducting Transmon qubit to a high-impedance $645\ \Omega$ microwave resonator. Doing so leads to a large qubit-resonator coupling rate $g$, measured through a large vacuum Rabi splitting of $2g\simeq 910$\ MHz. The coupling is a significant fraction of the qubit and resonator oscillation frequencies $\omega$, placing our system close to the ultra-strong coupling regime ($\bar{g}=g/\omega=0.071$ on resonance). Combining this setup with a vacuum-gap Transmon architecture shows the potential of reaching deep into the ultra-strong coupling $\bar{g} \sim 0.45$ with Transmon qubits.

\end{abstract}

\maketitle
\section{Introduction}
Cavity QED is a study of the light-matter interaction between atoms and the confined electro-magnetic field of a cavity~\cite{raimond_manipulating_2001}.
For an atom in resonance with the cavity, a single excitation coherently oscillates with vacuum Rabi frequency $g$ between the photonic and atomic degree of freedom if $g$ exceeds the rate at which excitations decay into the environment (the strong coupling condition).
Spectroscopically, this is observed as a mode-splitting (vacuum Rabi splitting) with distance $2g$. If the coupling is small with respect to the resonator (atomic) frequency $\omega_r$ ($\omega_a$), $g\ll \omega_r,\omega_a$, and the frequencies respect the condition $|\omega_a-\omega_r|\ll|\omega_r+\omega_a|$, the interaction is faithfully described with the Jaynes-Cummings (JC) model~\cite{jaynes_comparison_1963}.  
As the coupling becomes a considerable fraction of $\omega_r$ or $\omega_a$, typically $\bar{g}=g/\omega_{r,a} \sim 0.1$, the JC model no longer applies and the interaction is better described by the Rabi model~\cite{rabi_process_1936,braak_semi-classical_2016,arxiv_rossatto_spectral_2016,xie_quantum_2017}. 
This ultra-strong coupling (USC) regime shows the breakdown of excitation number conservation, however excitation parity remains conserved for arbitrarily large $\bar{g}$~\cite{casanova_deep_2010}. 
%
%
The key prediction for the deep-strong coupling (DSC) regime, where $\bar{g} \sim 1$,  is a symmetry breaking of the vacuum (i.\@ e.\@ qualitative change of the ground state) similar to the Higgs mechanism~\cite{garziano_vacuum-induced_2014}. The prospect of probing these new facets of light-matter interaction, in addition to potential applications in quantum information technologies~\cite{romero_ultrafast_2012,arxiv_stassi_quantum_2017}, has spurred many experimental efforts to reach increasingly large coupling rates. 
%

Experimentally, strong coupling has been achieved in systems with atoms~\cite{thompson_observation_1992,raimond_manipulating_2001,mckeever_experimental_2003,birnbaum_photon_2005}, and various solid-state implementations, including superconducting circuits with different types of qubits~\cite{wallraff_strong_2004,clarke_superconducting_2008,petersson_circuit_2012}, and semiconductor systems~\cite{reithmaier_strong_2004}. 
A different category of experiments using an ensemble of $N$ emitters benefit from a $\sqrt{N}$ enhancement of the coupling and therefore strong coupling has been observed in a wide variety of systems~\cite{herskind_realization_2009,putz_protecting_2014,zhu_coherent_2011}.
With such ensembles USC has been shown in the optical and THz frequency domain~\cite{anappara_signatures_2009,scalari_ultrastrong_2012,gambino_exploring_2014,schwartz_reversible_2011}. 
The only platform that observed higher coupling rates with a single emitter uses a superconducting circuit with a flux qubit. Pioneered by the experiments of Refs.~\cite{niemczyk_circuit_2010,forn-diaz_observation_2010}, experiments in the DSC regime have now been achieved with flux qubits coupled to resonators~\cite{yoshihara_superconducting_2017} as well as an electro-magnetic continuum~\cite{forn-diaz_ultrastrong_2016-1}. Additionally, the U/DSC coupling regime of the Rabi model was the subject of recent analog quantum simulations~\cite{arxiv_langford_experimentally_2016,arxiv_braumuller_analog_2016}. 
%

Here we explore coupling strengths at the edge of the USC regime in circuit QED using a superconducting Transmon qubit~\cite{koch_charge-insensitive_2007} coupled to a microwave cavity that has a high characteristic impedance. When the Transmon and fundamental mode of the cavity are resonant, we spectroscopically measure a coupling $g/2\pi=455$ MHz, corresponding to $\bar{g}=0.071$. With the prospect of maximizing the Transmon analogue of the dipole moment~\cite{bosman_2017_vacuum}, we show how this system could approach its theoretical upper limit~\cite{SI,jaako_ultrastrong-coupling_2016_comments}
\begin{equation}
	2g \lesssim \sqrt{\omega_r \omega_a\ }\ .
\end{equation}

As in previous implementations, the cavity is inherently a multi-mode system. In addition to this first deviation from the Rabi model, this architecture differentiates itself from previous implementations of an ultra-strong Rabi interaction by the weak anharmonicity of the Transmon. Its higher excitation levels become increasingly relevant with higher couplings and in the USC regime it cannot be considered a two-level system. The system studied here is therefore not a strict implementation of the Rabi model, but is still expected to bear many of the typical USC features and a proposal has been made to measure them~\cite{andersen_ultrastrong_2017}.

\section{Setup}
\begin{figure}[t!]
\includegraphics[width=0.47\textwidth]{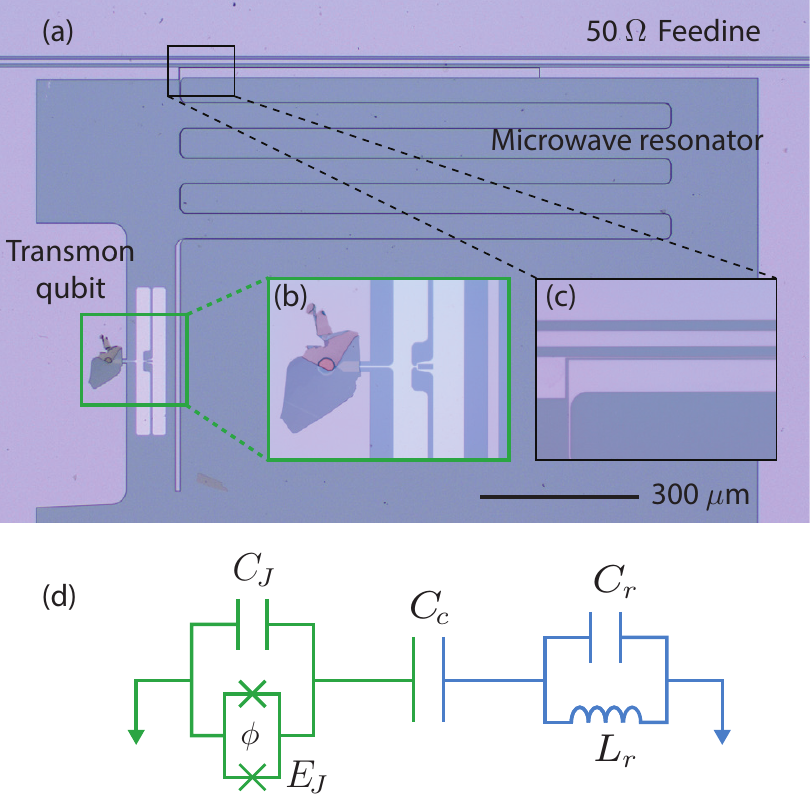}
\caption{(a) Optical image of the device, showing the $\lambda/2$-microwave resonator formed by a meandering $1 \ \mu$m wide stripline capacitively coupled to a $50\ \Omega$ feedline on one end and a Transmon qubit on the other end. (b) Zoom in on the Transmon qubit, showing the SQUID loop formed by two Josephson junctions and the parallel plate vacuum-gap capacitor formed by an aluminum qubit as bottom electrode and a multi-layer graphene as top electrode. (c) Zoom in on the capacitive coupler between the resonator and the feedline. (d) Equivalent lumped element circuit of the device, as derived in the supplementary information \cite{SI}.}
\label{fig:device}
\end{figure}

Our device, depicted in Fig.~\ref{fig:device}, consists of a high-impedance superconducting $\lambda/2$ microwave resonator~\cite{pozar_microwave_2009} capacitively over-coupled to a $50\ \Omega$ feedline on one end and coupled to a Transmon qubit on the other. The resonator is a $1\ \mu$m wide, $180$ nm thick and $\sim 6.5$ mm long meandering conductor. It is capacitively connected to a back ground plane through the $275\ \mu$m Silicon substrate as well as through vacuum/Silicon to the side ground planes.

The Transmon is in part coupled to ground through a vacuum gap capacitor, see Fig.~\ref{fig:device}(b). Its bottom electrode constitutes one island of the Transmon, the other plate is a suspended multi-layer graphene flake. The diameter of this capacitor is $15\ \mu$m with a gap of $150$ nm. This device was designed to couple the mechanical motion of the suspended multi-layer graphene to the Transmon qubit, where the coupling is mediated by a DC voltage offset~\cite{lahaye_nanomechanical_2009,pirkkalainen_cavity_2015}. Here we characterize the system at zero DC voltage, where the coupling to the motion is negligible. To enable tunability of the qubit frequency, a SQUID loop is incorporated such that the Josephson energy $E_J$ can be modified using an external magnetic field. $E_J(\phi)$ is a function of the flux through the SQUID loop $\phi$ following $E_J(\phi)=E_{J,\text{max}} \cos(\pi\phi /\phi_0)$ where $\phi_0=h/2e$ is the superconducting flux quantum.

We fabricate our devices in a three-step process. First we define our microwave resonators on a $275\ \mu$m Silicon substrate using reactive ion etching of molybdenum-rhenium alloy~\cite{singh_molybdenum-rhenium_2014}. Subsequently, $\mbox{Al/AlO}_x \mbox{/Al}$ Josephson junctions are fabricated using aluminum shadow evaporation~\cite{dolan_offset_1977}. Finally, we stamp a multi-layer graphene flake on the $15\mu$m diameter opening in the ground plane using deterministic dry viscoelastic stamping technique~\cite{castellanos-gomez_deterministic_2014}. From room temperature resistance measurements, optical and SEM images we observe that the flake is suspended, though folded, and that it does not short the qubit to ground.

This device implements the circuit shown in Fig.~\ref{fig:device}~\cite{SI}. Following circuit quantization~\cite{devoret__1997}, we find that the dynamics of the system are governed by the Hamiltonian
\begin{equation}
	\hat{H}=\hbar\omega_r \hat{a}^\dagger\hat{a}+\hbar\omega_a \hat{b}^\dagger\hat{b}-\frac{E_c}{2}\hat{b}^\dagger{b}^\dagger\hat{b}\hat{b}+\hbar g(\hat{a}+\hat{a}^\dagger)(\hat{b}+\hat{b}^\dagger)\ ,
	\label{eq:hamiltonian}
\end{equation}
where $\hat{a}$ ($\hat{b}$) is the annihilation operator for resonator (Transmon) excitations. The bare resonator and Transmon frequencies are given by $\omega_r = 1/\sqrt{L_rC_{r,\text{eff}}}$, $\omega_a = \sqrt{8E_JE_c}-E_c $, the charging energy is given by $E_c=e^2/2C_{J,\text{eff}}$ and the coupling strength 
\begin{equation}
	g\simeq\sqrt{\frac{\omega_a\omega_r}{4(1+\frac{C_J}{C_c})(1+\frac{C_r}{C_c})}}\ .
\end{equation}
The dependence on the flux $\phi$ is omitted in the expression of the coupling strength and Transmon frequency for clarity. It is important to distinguish the capacitances $C_c,C_J,C_r$ from the effective capacitances 
\begin{equation}
\begin{split}
	C_{J,\text{eff}} &= \frac{C_JC_r+C_JC_c+C_cC_r}{C_r+C_c}\ , \\
	C_{r,\text{eff}}&= \frac{C_JC_r+C_JC_c+C_cC_r}{C_J+C_c}\ .
\end{split}
\label{eq:eff_cap}
\end{equation}
The former correspond to the physical circuit elements, whereas the latter lead to the correct eigen-frequencies of the resonator and Transmon, defined as the oscillation rate of charges through the inductance $L_r$ and Josephson junction respectively.
Using finite-element simulation software, the qubit is designed such that its capacitance to ground $C_J = 51$ fF and its coupling capacitor is $C_c=9$ fF. The parameters of other circuit elements will be extracted from the data. We will denote the lowest three eigen-states of the Transmon by $\ket{g}$, $\ket{e}$ and $\ket{f}$ with increasing energies.

We characterize our device at a temperature of 15 mK, mounted in a radiation-tight box. From a vector network analyzer we send a microwave tone that is heavily attenuated before being launched on the feedline of the chip. 
The transmitted signal is send back to the vector network analyzer through a circulator and a low-noise HEMT amplifier. 
This setup is detailed in the supplementary information~\cite{SI}. It allows us to probe the absorption of our device and thus the energy spectrum of the Hamiltonian (\ref{eq:hamiltonian}).
At high driving power we measure the bare cavity resonance~\cite{bishop_response_2010} to have a total line-width of $\kappa=2\pi \times 29.3$ MHz and a coupling coefficient of $\eta=\kappa_\text{c}/\kappa=0.96$, giving the ratio between the coupling rate $\kappa_\text{c}$ and total dissipation rate $\kappa=\kappa_\text{c}+\kappa_\text{i}$, where $\kappa_\text{i}$ is the internal dissipation rate.
\section{Results}
\begin{figure}[b!]
\includegraphics[width=0.37\textwidth]{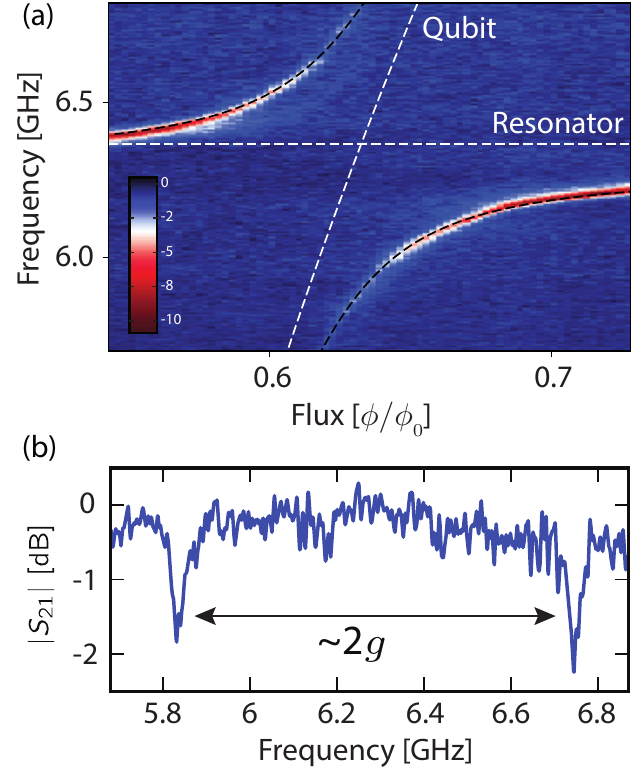}
\caption{\label{fig:wide} (a) Color plot of the feedline transmission $|S_{21}|$ as a function of magnetic field and frequency, showing the coupling between the resonator and qubit. The magnetic flux penetrating the SQUID loop incorporated in the Transmon qubit changes the Josephson energy, which results in the tunability of the $\ket{g}-\ket{e}$ transition energy of the qubit. Close to resonance the qubit and resonator show an avoided crossing centered at $6.23$ GHz. Black dashed lines correspond to fits to this data. The horizontal (oblique) white dashed line corresponds to the bare resonator (Transmon) frequency, $\omega_r/2\pi$ ($\omega_a(\phi)/2\pi$). The lack of symmetry around the crossing point is due to the flux dependence of the coupling strength $g$. (b) Trace of the microwave response where the qubit and cavity are close to resonance showing an anti-crossing of $2g\simeq2\pi \times 910$ MHz, and a line-width of $28$ MHz.}
\label{fig:single_tone}
\end{figure}
With a current biased coil, we can control the magnetic field and tune the effective $E_J$ to bring the qubit in resonance with the cavity. Where the Transmon and resonator frequencies cross, we measure a vacuum Rabi splitting which gives an estimate of the coupling rate $2g/2\pi \simeq 910$~MHz as shown in Fig.~\ref{fig:single_tone}.

\begin{figure}
\includegraphics[width=0.37\textwidth]{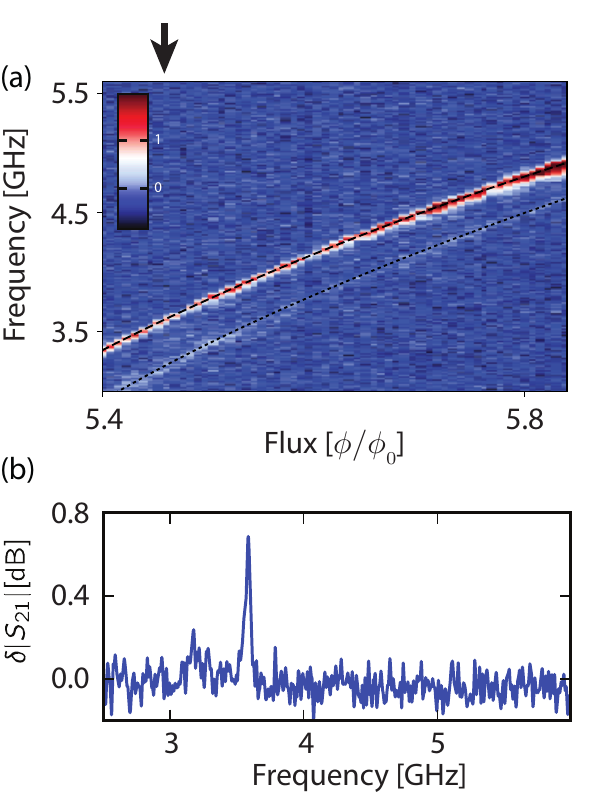}
\caption{(a) Color plot of the change in transmission ($\delta|S_{21}|$) of a microwave tone at resonance with the cavity at single-photon power as a function of frequency of a secondary qubit-drive tone and magnetic field. Due to the qubit-state dependent frequency shift of the cavity, the transmission changes as the drive tone excites the qubit, tracing the qubit $\ket{g}-\ket{e}$ transition frequency as a function of magnetic field. The secondary faint resonance corresponds to the $\ket{e}-\ket{f}$ transition of the thermally excited population in state $\ket{e}$. Power broadening is visible when the qubit transition frequency is close to the cavity. The power of the drive tone delivered to the feedline is constant, but close to the cavity a larger portion of the power is delivered to the qubit. Dashed lines correspond to fits to the data. (b) Line-cut of the color plot where the qubit frequency is at $\omega_a=2\pi \times 3.586$ GHz with a line-width of $38$ MHz. The second peak corresponds to the first to second exited state transition, which corresponds to the (dressed) anharmonicity, a rough approximation of $E_c/h= 370$ MHz.}
\label{fig:two_tone}
\end{figure}
In Fig.~\ref{fig:two_tone}, we show the result of performing two-tone spectroscopy to probe the qubit frequency~\cite{blais_cavity_2004,wallraff_approaching_2005}. When the qubit is detuned from the cavity, the resonator acquires a frequency shift which is dependent on the state of the qubit. Hence, probing the transmission of the feedline at the cavity resonance (shifted by the qubit in the ground state), while exciting the qubit with another microwave tone, will cause the transmission to change by a value $\delta|S_{21}|$ due to the qubit-state dependent shift. In Fig.~\ref{fig:two_tone}(a) we measure the spectral response of the qubit for different magnetic fields. As the magnetic flux through the SQUID loop tunes the qubit frequency we track the ground to first exited state transition as a function of magnetic field. Since the probe power is kept constant during this experiment a clear power broadening of the qubit is visible, because more of the power is delivered to the qubit as it is closer to the cavity in frequency. The $38$ MHz line-width of this resonance translates to very short coherence times ($T_1\sim40$~ns) compared to typical implementations \cite{houck_controlling_2008}. Purcell losses contribute less than $2$ MHz to this line-width and the full origin of this high dissipation remains unknown. The secondary faint resonance corresponds to the spectral response of the first to second exited state transition of the Transmon due to some residual occupation of the first excited state. The difference in frequency between both transitions provides an estimate of the charging energy (or equivalently the anharmonicity of the Transmon), $E_c/h \sim 370$ MHz. In reality what we measure is a quantity that is dressed by the interaction with the cavity and diverges from the bare value of $E_C$.


We fit a numerical diagonalization of the Hamiltonian detailed in the supplementary information~\cite{SI} to the acquired data, obtaining the fits shown as dashed lines in Figs.~\ref{fig:single_tone},\ref{fig:two_tone}. We thereby obtain the Hamiltonian parameters $E_c/h = 300$ MHz, $g/2\pi=455$ MHz (on resonance), $\omega_r/2\pi=6.367$ GHz and $E_{J,\text{max}}/h=46$ GHz. Combined with our knowledge of the capacitances $C_J$ and $C_c$, we extract the following values for the circuit elements of the resonator: $C_r=57.1$ fF and $L_r=9.65$ nH. If we assume that the parallel LC oscillator corresponds to the fundamental mode of a $\lambda/2$ resonator, then the resonators effective impedance $Z_r=\sqrt{L_r/C_r}=411\ \Omega$ is related to the characteristic impedance of the transmission line through $Z_0=\pi Z_r/2$~\cite{SI}, yielding a value $Z_0=645\ \Omega$. \\

\section{Towards higher coupling in Transmon systems: a proposal}

In the circuit of Fig.~\ref{fig:device}(d), the coupling rate is limited following
\begin{equation}
	\frac{g}{\sqrt{\omega_a\omega_r}} = \frac{1}{2}\sqrt{\frac{1}{1+\frac{C_J}{C_c}}\frac{1}{1+\frac{C_r}{C_c}}}\leq \frac{1}{2}\ .
	\label{eq:coupling}
\end{equation}
The highest couplings are therefore achieved by maximizing two capacitance ratios: $C_c/C_J$ and $C_c/C_r$. In the language of cavity QED with natural atoms, maximizing the first ratio is equivalent to increasing the dipole moment of the atom which is done by using Rydberg atoms~\cite{raimond_manipulating_2001}. Maximizing the second ratio increases the vacuum fluctuations of the cavities electric field as performed in alkali-atom experiments in a very small optical cavity~\cite{thompson_observation_1992}.

In the regime $C_c/C_J, C_c/C_r \gg 1$ the effective capacitances of Eq.~(\ref{eq:eff_cap}) are approximated by
\begin{equation}
	C_{J,\text{eff}} = C_{r,\text{eff}} \simeq C_J+C_r\ ,
\end{equation}
these capacitances being the quantities to minimize to increase the coupling. Maximizing the coupling whilst keeping the resonator and Transmon frequencies constant therefore requires a large increase in the inductances. In other words, the higher the impedance of the resonator and the higher the ratio $E_c/E_J$ in the Transmon, the higher the coupling.

The highest ratio of $E_c/E_J$ for which we remain in the Transmon regime is $\sim 1/20$~\cite{koch_charge-insensitive_2007}. Combined with a typical choice of the Transmon frequency $\omega_a/2\pi=6$ GHz, compatible with most microwave experimental setups, we obtain a value of the Transmons total capacitance $C_{J,\text{eff}} = C_{r,\text{eff}} \simeq C_J+C_r\simeq20$ fF. Choosing $C_J=C_r=10$ fF maximizes Eq.~\ref{eq:coupling}. Fixing the resonator frequency to $\omega_r=\omega_a$ leads to a value for resonators characteristic impedance: $722\ \Omega$ if a $\lambda/4$ resonator is used, $1.44$ k$\Omega$ for a $\lambda/2$ resonator and $918\ \Omega$ for a lumped element resonator. The coupling achieved now depends on the value of the coupling capacitor. For $C_c=200$ fF for example, $\bar{g}=0.45$ and the system is deep in the USC regime.

\begin{figure}
\includegraphics[width=0.5\textwidth]{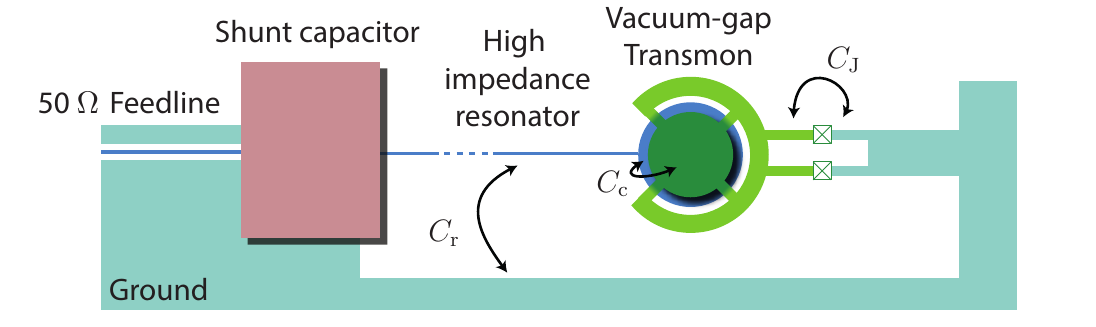}
\caption{Proposal for larger couplings which combines the use of a $\lambda/4$ high impedance resonator and a vacuum gap Transmon. The $\lambda/4$ can be coupled to a $50\ \Omega$ feedine through the use of a shunt capacitor~\cite{bosman_broadband_2015}. For $C_J=C_r=10$ fF and $C_c=200$ fF the system is expected to reach far into the USC regime $\bar{g}=0.45$ on resonance.}
\label{fig:proposal}
\end{figure}

In Ref.~\cite{bosman_2017_vacuum}, the USC regime was reached by increasing the first capacitance ratio of Eq.~(\ref{eq:coupling}), $C_c/(C_J+C_c)\simeq 0.9$ through the use of a vacuum-gap coupling capacitor. In this work, the large coupling is reached by increasing the second ratio $C_c/(C_r+C_c)\simeq 0.136$ above usual values through the use of a high impedance resonator whilst the first capacitance ratio remains modest $C_J/(C_J+C_c)\simeq 0.15$. Combining both approaches into a single device represented schematically in Fig.~\ref{fig:proposal} would allow experimentally, reaching deep into the USC regime $\bar{g}=0.45$ by matching the circuit parameters presented previously. The values $C_J=10$ fF and $C_c=200$ fF can be easily achieved experimentally reproducing the vacuum-gap Transmon architecture of Ref.~\cite{bosman_2017_vacuum} with a smaller gap and larger capacitive plate, maybe even by replacing the vacuum gap by a dielectric. The use of a $\lambda/4$ resonator rather than a $\lambda/2$ is preferable as it decreases the impedance needed as well as increases the frequency spacing between the fundamental and higher modes. Moving to a $\lambda/4$ resonator makes the current architecture sufficient in terms of resonator impedance. The impedance could be further increased using a high kinetic inductance based resonator~\cite{samkharadze_high-kinetic-inductance_2016} or by using an array of Josephson junctions~\cite{masluk_microwave_2012}. 

This proposal is however limited by the underlying assumption that only a single mode of the resonator participates in the dynamics of the system. However for larger coupling rates, the higher modes no longer play a weak perturbative role \cite{arxiv_gely_divergence-free_2017}. Exploring the exact consequences of this fact on the observable USC phenomena that can be observed is outside the scope of this work, as is determining alternatives to probing the system spectroscopically to show for example the non-trivial ground state that one would expect in this regime. For a detailed study of these topics, we refer the reader to Ref.~\cite{andersen_ultrastrong_2017}.

\section{Conclusion}

We have shown that it is possible to enhance the coupling between a microwave resonator and a Transmon qubit by increasing the impedance of the resonator to $645\ \Omega$ compared to typical $50\ \Omega$ implementations. In doing this we reach a coupling rate of $g/2\pi=455$ MHz at resonance, which is close to the ultra-strong coupling regime ($\bar{g}=0.071$). We have shown that by optimizing this strategy through sources of high inductance, combined with a vacuum-gap Transmon architecture, we have the potential of reaching far into the ultra-strong coupling regime. \\

\textit{Acknowledgments} The authors thank Alessandro Bruno, Leo DiCarlo, Nathan Langford, Adrian Parra-Rodriguez and Marios Kounalakis for useful discussions.

\bibliography{../../../projects/mario/library.bib,extra,arxiv}

\end{document}



\title{Supplementary Material: Approaching ultra-strong coupling in Transmon circuit-QED using a high-impedance resonator}

\author{Sal J. Bosman}
\affiliation{%
Kavli Institute of NanoScience, Delft University of Technology,\\
PO Box 5046, 2600 GA, Delft, The Netherlands.
}%

\author{Mario F. Gely}
\affiliation{%
Kavli Institute of NanoScience, Delft University of Technology,\\
PO Box 5046, 2600 GA, Delft, The Netherlands.
}%

\author{Vibhor Singh}%
 \affiliation{%
Department of Physics, Indian Institute of Science, Bangalore 560012, India
}%

\author{Daniel Bothner}%
 \affiliation{%
Kavli Institute of NanoScience, Delft University of Technology,\\
PO Box 5046, 2600 GA, Delft, The Netherlands.
}%

\author{Andres Castellanos-Gomez}
\affiliation{%
 Instituto de Ciencia de Materiales de Madrid, CSIC, Madrid 28049, Spain}

\author{Gary A. Steele}

\affiliation{%
Kavli Institute of NanoScience, Delft University of Technology,\\
PO Box 5046, 2600 GA, Delft, The Netherlands.
}%
\date{\today}

\maketitle
\onecolumngrid
\title{Supplementary material}

\maketitle

\section{Experimental setup}

\begin{figure}[h!]
\includegraphics[width=0.3\textwidth]{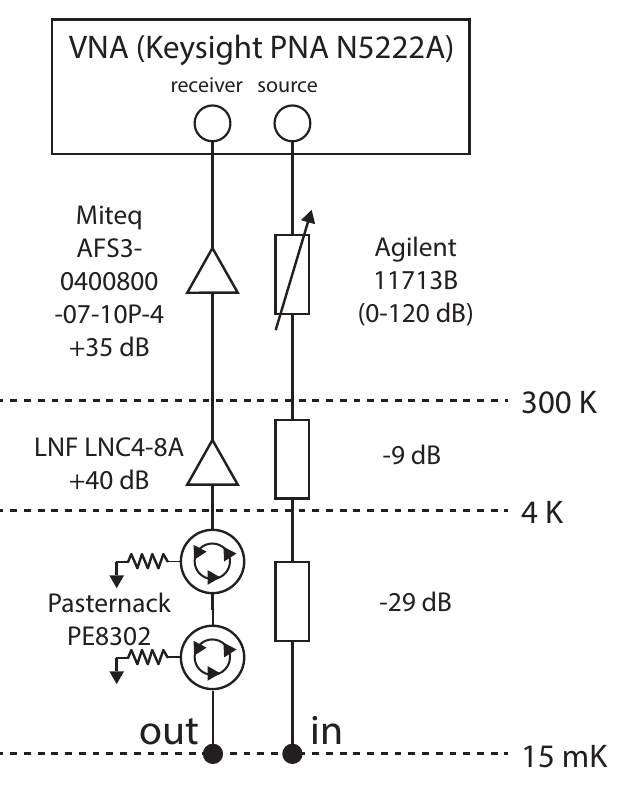}
\caption{Schematic representation of the experimental setup described in the main text. The input microwave line is heavily attenuated down to the 15 mK stage where the sample is placed. On the return line the sample is isolated from the noise of the amplification stage through the use of two circulators. A HEMT amplifier is placed at the 4 K stage and a room temperature amplifier completes the return line.}
\label{fig:setup}
\end{figure}

A schematical representation of our experimental setup is presented in Fig~\ref{fig:setup}.

\section{Equivalent circuit}
\begin{figure}[b]
\includegraphics[width=\textwidth]{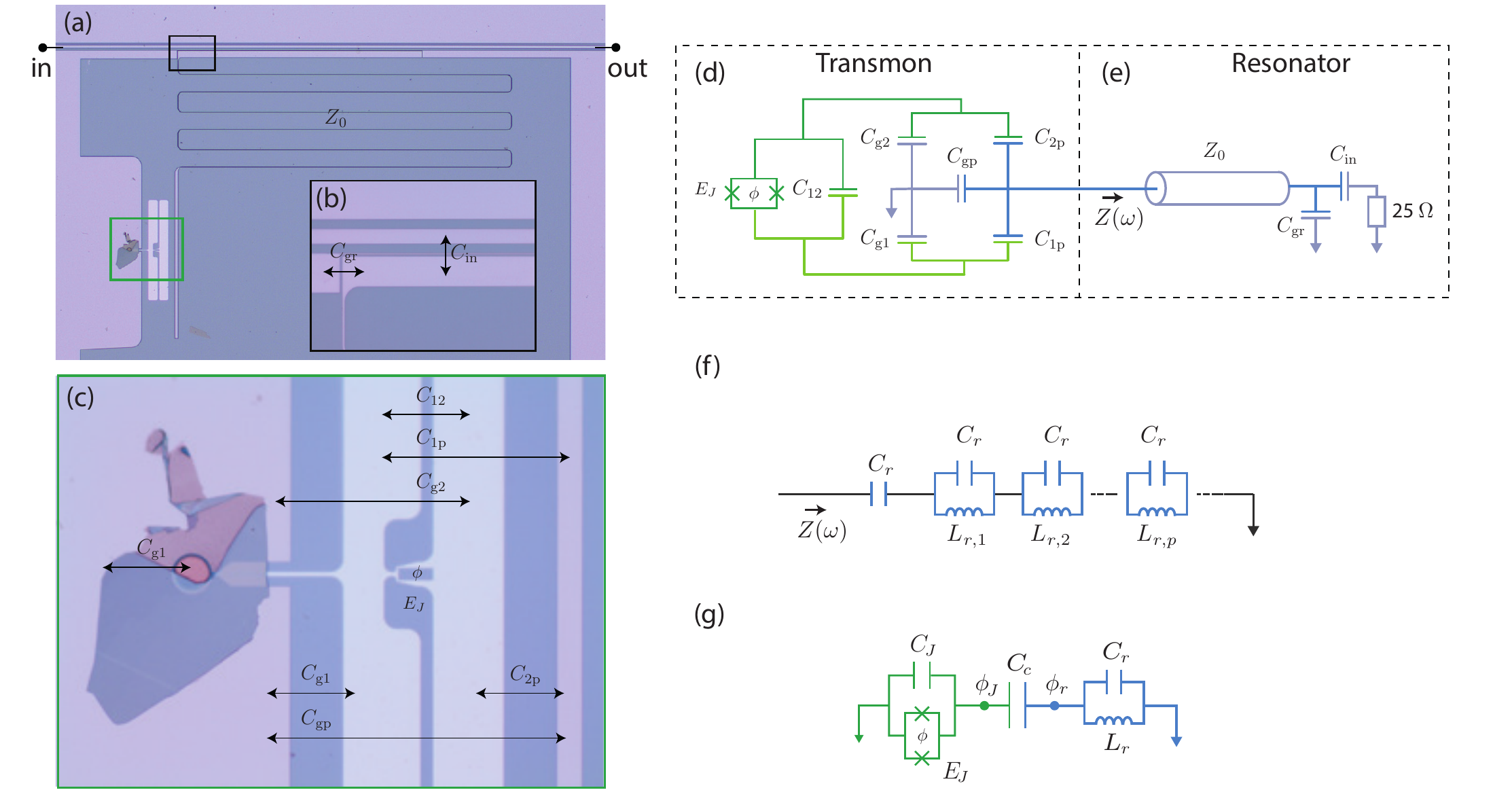}
\caption{(a-c) Optical image of the device annotated with the various circuit parameters used in the derivation of the system Hamiltonian. (d,e) Schematic representation of the devices electrical circuit. (f) Foster equivalent circuit of the (lossless) resonator. The left-most LC oscillator corresponds to the fundamental mode of the resonator and is the only resonance considered in our analysis. (g) Simplified circuit representation of the device.}
\label{fig:circuit}
\end{figure}

The capacitance network of the Transmon and its coupling to the resonator shown in Fig.~\ref{fig:circuit}(c,d) can be reduced to using two capacitances $C_J,C_c$ as shown in Fig.~\ref{fig:circuit}(g) following the methodology of Ref.~\cite{koch_charge-insensitive_2007}. Using the finite element solver ANSYS Maxwell, we compute the value of each of the physical capacitances. This results in the values $C_J = 51$ fF, $C_c=9$ fF.\\

The input impedance $Z(\omega)$ seen by the Transmon is that of a finite section of a transmission line, or resonator, with characteristic impedance $Z_0$, terminated by a capacitive connection $C_s=C_{\text{in}}+C_{\text{gr}}$ to ground (we ignore the $25\ \Omega$ load that constitutes the feedline along with all other sources of dissipation). If this capacitance is small (\textit{i.e.} $\omega Z_0 C_s \ll 1$ with $\omega$ in the GHz range), the resonator is a $\lambda/2$ cavity with $Z(\omega) = -iZ_0\cot(\pi\omega/\omega_r)$ where $\omega_r$ would be the resonance frequency of the fundamental mode.\\ 

Fosters reactance theorem states that these impedances can be synthesized by a series combination of LCR oscillators \cite{foster_reactance_1924,nigg_black-box_2012-1}, each corresponding to a mode of the resonator (see Fig.~\ref{fig:circuit}(f)). The absence of dissipation sources means that each oscillator is fully characterized by (i) its angular frequency $\omega_{r,p}$ defined as the real parts of the zeros of the admittance $Y(\omega)=Z(\omega)^{-1}$ (such that $\omega_{r,p} = \text{Re}(\tilde{\omega}_{r,p})$, $Y(\tilde{\omega}_{r,p})=0$) and (ii) by its impedance $Z_{r,p} = 2/(\omega_{r,p}\text{Im}Y'(\omega_{r,p}))$. The input impedance of a $\lambda/2$ resonator is then equivalent to that of a capacitance $C_r = \pi/2\omega_0 Z_0$ in series with an infinite series of LC oscillators as shown in Fig.~\ref{fig:circuit}(f). The impedance and frequency, as well as the capacitance and inductance of each LC oscillator, is given by:
\begin{equation}
  \begin{split}
    \omega_{r,p}    &= p\omega_r      \ ,\\
Z_{r,p}  &= \displaystyle\frac{2Z_0}{p\pi}    \ ,\\
C_{r,p}&= C_r=\displaystyle\frac{\pi}{2\omega_r Z_0}      \ ,\\
L_{r,p}  &= \displaystyle\frac{1}{p^2}\frac{2Z_0}{\pi \omega_r}    \ ,
  \end{split}
\end{equation}
with $p=1,2,3,...$\\

Although the capacitance of the resonator on the coupler end is substantial, warranting the name of over-coupled $\lambda/2$ resonator, the decision to approximate it as an ideal $\lambda/2$ can be verified by calculating the phase velocity $v_\varphi=\lambda_1 /2\pi\omega_r$ in the line. The wavelength of the fundamental mode $\lambda_1$ is given by twice the total length of the line $\lambda_1=2\times 7.4$ mm (including the length of the coupling pads), yielding a phase velocity of $\sim1\times10^8$ m/s. Approximating the distance from the conductor to side ground plane to a constant $\sim1$ mm (the dominating capacitance coming from the back plane), we calculate an effective relative dielectric constant of $\epsilon_r\simeq6.9$ \cite{wadell_transmission_1991-1} corresponding to a theoretical phase velocity $c/\sqrt{\epsilon_r}\simeq1.14\times10^8\ \text{m.s}^{-1}$. This value is a bit larger than the measured speed as expected from over-coupling our resonator. The impedance extracted from our fits is then a slight over-estimate.\\

The series capacitance $C_r = \pi/2\omega_0 Z_0\simeq57$ fF dominates over the coupling capacitor $C_c=9$ fF, and therefore can be ignored. As established using QUCS~\cite{brinson_qucs:_2009}, the second mode of the resonator shifts the spectrum by $~5$ MHz and the third by less than $1$ MHz. These shifts being smaller than the measured linewidths, we only retain a single LC oscillator of this lumped element decomposition, with capacitance $C_r = C_{r,1}$ and inductance $L_r=L_{r,1}$.

\section{Hamiltonian derivation}

The derivation of the Hamiltonian is described in great detail in the supplementary material of Ref.~\cite{arxiv_gely_divergence-free_2017} or in Refs.~\cite{jaako_ultrastrong-coupling_2016,andersen_ultrastrong_2017}. Using the tools of circuit quantization \cite{devoret__1997}, we reach the following Hamiltonian
\begin{equation}
    \begin{split}
       \hat{H}_{\text{cir}} &=\frac{C_c+C_r }{2 C_*^2}\hat{q}_J^2-E_J(\phi)\cos\bigg(\pi\frac{\hat{\phi}_J}{\phi_0}\bigg)\\
         &+\frac{C_c+C_J}{2 C_*^2}\hat{q}_r^2+ \frac{\hat{\phi}_{r}^2}{2L_r}\\
         &+\frac{C_c}{C_*^2}\hat{q}_J \hat{q}_r\ ,
    \end{split}
\end{equation}
where we introduced the capacitance $C_*$ defined by
\begin{equation}
    C_*^2 = C_c C_J + C_cC_r+ C_JC_r\ .
\end{equation}
The node fluxes $\hat{\phi}_{i}$, with $i=J$ or $i=r$, shown in Fig.~\ref{fig:circuit}(g) are defined classically from the voltage of the node $V_{i}$ as
\begin{equation}
  \phi_{i}(t)=\int_{-\infty}^{t}v_{i}(t')dt'\ ,
\end{equation}
and the charges $\hat{q}_{i}$ are their canonical conjugate variables such that the usual commutation relation is observed
\begin{equation}
  [\hat{\phi}_i,\hat{q}_j] = i\hbar\delta_{ij}\ .
\end{equation}
$\phi_0=h/2e$ is the superconducting flux quantum. We define the creation and annihilation operators for the photonic degree of freedom with frequency $\omega_r = 1/\sqrt{L_rC_{r,\text{eff}}}$ of the resonator as

\begin{equation}
  \hat{\phi}_r = -i\sqrt{\frac{\hbar}{2}\sqrt{\frac{L_r}{C_{r,\text{eff}}}}}(\hat{a}-\hat{a}^\dagger)\ ,
\end{equation}
\begin{equation}
  \hat{q}_r = \sqrt{\frac{\hbar}{2}\sqrt\frac{C_{r,\text{eff}}}{L_r}}(\hat{a}+\hat{a}^\dagger)\ ,
\end{equation}
where $C_{r,\text{eff}} = C_*^2/( C_c+C_J)$ is the capacitance of the resonator modified by the presence of the Transmons capacitances. We introduce the eigen-states $\{\ket{k}\}_k$ and corresponding eigen-energies $\hbar\epsilon_k$ (the state $0$ will be taken as reference $\epsilon_0=0$) which diagonalize the Cooper pair box Hamiltonian
\begin{equation}
\begin{split}
    \hat{H}_{\text{CPB}}&=\frac{C_c+C_r }{2 C_*^2}\hat{q}_J^2-E_J(\phi)\cos\bigg(\pi\frac{\hat{\phi}_J}{\phi_0}\bigg) \\
    &= 4E_c\hat{n}_J^2-E_J(\phi)\cos(\hat{\delta})\ ,
\end{split}
\end{equation}
where $\hat{n}_J=\hat{q_J}/2e$ is the quantum number of Cooper-pairs on the island conjugate to $\hat{\delta}=\pi\hat{\phi}_J/\phi_0$ the superconducting phase difference across the junction, and $E_c=e^2/2C_{J,\text{eff}}$ is the charging energy of the island. $C_{J,\text{eff}}= C_*^2/(C_c+C_r)$ is the capacitance of the Transmon modified by the presence of the resonator capacitances. We will purposefully omit the dependence of all eigen-states and eigen-energies on $\phi$ to simplify notation. This notably yields the following expression for $\hat{q}_J$
\begin{equation}
    \hat{q}_J = 2e\sum_{k,l\geq 0}\bra{k}\hat{n}_J\ket{l}\ket{k}\bra{l}\ .
\end{equation}
Omitting constant energy contributions, the Hamiltonian can now be written
\begin{equation}
    \begin{split}
         \hat{H}/\hbar&=\sum_{i>0}\epsilon_k\ket{k}\bra{k} + \omega_r\hat{a}^\dagger\hat{a}+ \sum_{k,l\geq0}g_{kl}(\hat{a}+\hat{a}^\dagger) \ket{k}\bra{l}\ ,
    \end{split}
    \label{eq:hamiltonian_num}
\end{equation}
where we introduce the coupling rates 
\begin{equation}
g_{ij} = \frac{2e}{\hbar}\frac{C_c}{C_*^2}\sqrt{\frac{\hbar}{2}\sqrt\frac{C_{r,\text{eff}}}{L_r}}\bra{k}\hat{n}_J\ket{l}\ .
\end{equation}
An additional approximation can be made by assuming that the ratio $E_J/E_c>20$ is high enough to place the Cooper pair box in the Transmon regime. The Transmon can then be modeled as a Duffing oscillator \cite{andersen_ultrastrong_2017} such that the CPB Hamiltonian reduces to 
\begin{equation}
  \hat{H}_{\text{CPB}} \simeq \hbar\omega_a \hat{b}^\dagger\hat{b}-\frac{E_c}{2}\hat{b}^\dagger{b}^\dagger\hat{b}\hat{b}\ ,
\end{equation}
where $\omega_a = \sqrt{8E_J(\phi)E_c}-E_c $ and we introduced 
\begin{equation}
  \begin{split}
    \hat{\phi}_J&=-i\phi_0\left(\frac{2E_c}{E_J}\right)^{1/4}(\hat{b}-\hat{b}^\dagger)\ ,\\
    \hat{n}_J&=\left(\frac{E_J}{32E_c}\right)^{1/4}(\hat{b}+\hat{b}^\dagger)\ .
  \end{split}
\end{equation}
This simplifies the Hamiltonian to
\begin{equation}
  \hat{H}=\hbar\omega_r \hat{a}^\dagger\hat{a}+\hbar\omega_a \hat{b}^\dagger\hat{b}-\frac{E_c}{2}\hat{b}^\dagger{b}^\dagger\hat{b}\hat{b}+\hbar g(\hat{a}+\hat{a}^\dagger)(\hat{b}+\hat{b}^\dagger)\ .
  \label{eq:hamiltonian}
\end{equation}
While the above form is useful in simplifying the analytical expression of the coupling, all numerical calculations presented in the main text are based on the exact diagonalization of Eq.~\ref{eq:hamiltonian_num}. The coupling can be cast into several forms:
\begin{equation}
  \begin{split}
    \hbar g&=2e\frac{C_c}{C_*^2}\left(\frac{E_J}{32E_c}\right)^{1/4}\sqrt{\frac{\hbar}{2}\sqrt\frac{C_{r,\text{eff}}}{L_r}}\\
    &=\frac{\hbar}{2}\frac{C_c}{C_*^2}\sqrt{\frac{1}{Z_{r,\text{eff}}Z_{a,\text{eff}}}}\\
    &\simeq\frac{\hbar}{2}\sqrt{\frac{\omega_a\omega_r}{(1+\frac{C_J}{C_c})(1+\frac{C_r}{C_c})}}\ .
  \end{split}
\end{equation}
Here we introduced the effective impedances of the two oscillators $Z_{i,\text{eff}}=\sqrt{L_i/C_{i,\text{eff}}}$, with $i=J,r$, where $L_J=\phi_0^2/4\pi^2E_J$. The last form is used to derive the maximum coupling architecture and the approximation made to derive it is $\omega_a=\sqrt{8E_JE_c}-E_c\simeq\sqrt{8E_JE_c}=1/\sqrt{L_aC_{J,\text{eff}}}$

\bibliography{../../../projects/mario/library.bib,extra_SI,arxiv}